\documentclass[preprint2]{aastex}

\usepackage{hyperref}
\usepackage{graphicx}

\newcommand{\taro}{K2-114} 
\newcommand{\tart}{K2-115}
\newcommand{\tarob}{K2-114b} 
\newcommand{\tartb}{K2-115b}

\newcommand{\rsun}{\ensuremath{R_\sun}}
\newcommand{\msun}{\ensuremath{M_\sun}}

\newcommand{\rstar}{\ensuremath{R_s}}
\newcommand{\mstar}{\ensuremath{M_s}}

\newcommand{\rhostar}{\ensuremath{\rho_s}}

\newcommand{\mearth}{\ensuremath{M_\earth}}

\newcommand{\rpl}{\ensuremath{R_{p}}}
\newcommand{\mpl}{\ensuremath{M_{p}}}

\newcommand{\rjup}{\ensuremath{R_{\rm J}}}
\newcommand{\mjup}{\ensuremath{M_{\rm J}}}

\newcommand{\teff}{\ensuremath{T_{\rm eff}}}
\newcommand{\teq}{\ensuremath{T_{\rm eq}}}
\newcommand{\logg}{\ensuremath{\log g_s}}
\newcommand{\feh}{[Fe/H]}

\newcommand{\vsini}{\ensuremath{V\sin(I)}}

\newcommand{\kms}{km~s$^{-1}$}
\newcommand{\ms}{m~s$^{-1}$}
\newcommand{\ergscm}{erg s$^{-1}$ cm$^{-2}$}

\newcommand{\kt}{{\it K2}}

\newcommand{\kp}{$K_{p}$}

\newcommand{\mic}{\ensuremath {\rm \mu m}}

\newcommand{\sig}[1]{\ensuremath{#1\sigma}}

\newcommand{\figr}[1]{Fig.~\ref{fig:#1}}

\newcommand{\secr}[1]{Sec.~\ref{sec:#1}}

\newcommand{\tabr}[1]{\mbox{Table~\ref{tab:#1}}}

\slugcomment{Submitted to AAS Journals}

\shorttitle{Two Transiting Warm Jupiters}
\shortauthors{Shporer et al.}

\begin{document}

\title{\tarob\ and \tartb: Two Transiting Warm Jupiters}

\author{
Avi Shporer\altaffilmark{1}, 
George Zhou\altaffilmark{2},
Benjamin J.~Fulton\altaffilmark{3,4,24},
Andrew Vanderburg\altaffilmark{2,24},
Nestor Espinoza\altaffilmark{5,6},
Karen Collins\altaffilmark{7},
David Ciardi\altaffilmark{8},
Daniel Bayliss\altaffilmark{9},
James D.~Armstrong\altaffilmark{10},
Joao Bento\altaffilmark{11},
Francois Bouchy\altaffilmark{9},
William D.~Cochran\altaffilmark{12},
Andrew Collier Cameron\altaffilmark{13},
Knicole Col\'on\altaffilmark{14},
Ian Crossfield\altaffilmark{15},
Diana Dragomir\altaffilmark{16,25},
Andrew W.~Howard\altaffilmark{4},
Steve B.~Howell\altaffilmark{17},
Howard Isaacson\altaffilmark{18},
John F.~Kielkopf\altaffilmark{19}, 
Felipe Murgas\altaffilmark{20,21}, 
Ramotholo Sefako\altaffilmark{22},
Evan Sinukoff\altaffilmark{3,4},
Robert Siverd\altaffilmark{23},
Stephane Udry\altaffilmark{9}\\
}

\altaffiltext{1}{Division of Geological and Planetary Sciences, California Institute of Technology, Pasadena, CA 91125, USA}
\altaffiltext{2}{Harvard–-Smithsonian Center for Astrophysics, 60 Garden Street, Cambridge, MA 02138, USA}
\altaffiltext{3}{Institute for Astronomy, University of Hawaii, Honolulu, HI 96822, USA}
\altaffiltext{4}{California Institute of Technology, Pasadena, CA, USA}
\altaffiltext{5}{Instituto de Astrof\'isica, Facultad de F\'isica, Pontificia Universidad Cat\'{o}lica de Chile, Av.~Vicu\~na Mackenna 4860,
782-0436 Macul, Santiago, Chile.}
\altaffiltext{6}{Millennium Institute of Astrophysics, Av.~Vicu\~na Mackenna 4860, 782-0436 Macul, Santiago, Chile.}
\altaffiltext{7}{Department of Physics \& Astronomy, Vanderbilt University, Nashville, TN 37235, USA}
\altaffiltext{8}{NASA Exoplanet Science Institute/Caltech, Pasadena, CA, USA}
\altaffiltext{9}{Observatoire Astronomique de l'Universit'e de Geneve, 51 ch.~des Maillettes, 1290 Versoix, Switzerland;}
\altaffiltext{10}{University of Hawaii Institute for Astronomy 34 Ohia Ku Street, Pukalani, HI 96768}
\altaffiltext{11}{Research School of Astronomy and Astrophysics, Mount Stromlo Observatory, Australian National University, Weston, ACT 2611, Australia.}
\altaffiltext{12}{McDonald Observatory and Department of Astronomy, University of Texas, Austin, TX 78812, USA}
\altaffiltext{13}{Centre for Exoplanet Science, SUPA School of Physics \& Astronomy, University of St Andrews, North Haugh, St Andrews KY16 9SS, UK}
\altaffiltext{14}{NASA Goddard Space Flight Center, Greenbelt, MD 20771, USA}
\altaffiltext{15}{Astronomy and Astrophysics Department, UC Santa Cruz, CA, USA}
\altaffiltext{16}{Massachusetts Institute of Technology, Cambridge, MA 02139 USA}
\altaffiltext{17}{NASA Ames Research Center, Moffett Field, CA 94035, USA}
\altaffiltext{18}{Astronomy Department, University of California, Berkeley, CA, USA}
\altaffiltext{19}{Department of Physics \& Astronomy, University of Louisville, Louisville, KY 40292, USA}
\altaffiltext{20}{Instituto de Astrof\'isica de Canarias, V\'ia L\'actea s/n, E-38205 La Laguna, Tenerife, Spain}
\altaffiltext{21}{Departamento de Astrof\'isica, Universidad de La Laguna, Spain}
\altaffiltext{22}{South African Astronomical Observatory, PO Box 9, Observatory, 7935, South Africa}
\altaffiltext{23}{Las Cumbres Observatory Global Telescope Network, 6740 Cortona Dr., Suite 102, Santa Barbara, CA 93117, USA}
\altaffiltext{24}{National Science Foundation Graduate Research Fellow}
\altaffiltext{25}{NASA Hubble Fellow}

%===============================================================================
\begin{abstract}

We report the first results from a search for transiting warm Jupiter exoplanets --- gas giant planets receiving stellar irradiation below about 10$^8$ \ergscm, equivalent to orbital periods beyond about 10 days around Sun-like stars.
We have discovered two transiting warm Jupiter exoplanets initially identified as transiting candidates in \kt\ photometry. 
\tarob\ has a mass of $1.85^{+0.23}_{-0.22}\ \mjup$, a radius of $0.942^{+0.032}_{-0.020}\ \rjup$, and an orbital period of 11.4~days.
\tartb\ has a mass of $0.84^{+0.18}_{-0.20}\ \mjup$, a radius of $1.115^{+0.057}_{-0.061}\ \rjup$, and an orbital period of 20.3~days.
Both planets are among the longest period transiting gas giant planets with a measured mass, and they are orbiting relatively old host stars. Both planets are not inflated as their radii are consistent with theoretical expectations. Their position in the planet radius - stellar irradiation diagram is consistent with the scenario where the radius - irradiation correlation levels off below about 10$^8$ \ergscm, suggesting that for warm Jupiters the stellar irradiation does not play a significant role in determining the planet radius. We also report our identification of another \kt\ transiting warm Jupiter candidate, EPIC 212504617, as a false positive.

\end{abstract}
%===============================================================================

\keywords{planetary systems: individual (\taro, \tart)}

%===============================================================================
\section{Introduction}
\label{sec:intro}
%===============================================================================

The number of known transiting hot Jupiter exoplanets --- gas giant exoplanets in short orbital periods of only a few days --- is now at the few hundreds. Despite that large number there are still several open questions related to this class of exoplanets. Many, if not most, of the known hot Jupiters have larger radii than theoretically expected \citep[e.g.,][]{weiss13, baraffe14, lopez16}. Although various explanations have been proposed none have completely solved this puzzle, suggesting there is more than a single mechanism at play here, and/or that we are missing some physics shaping planetary structure \citep{baraffe14}. Another hot Jupiter mystery is their formation and orbital evolution. While several theories have been put forth \citep[e.g.,][]{lin96, rasio96, alibert05, fabrycky07} it is still not clear how gas giant planets reach short orbits around Sun-like stars, at only a few 0.01~au.

One way to shed light on the above questions is to examine the population of {\it warm Jupiters} --- gas giant planets receiving stellar irradiation below 10$^8$ erg s$^{-1}$ cm$^{-2}$, corresponding to orbital periods beyond $\approx$10 days around Sun-like stars. Specifically, we would like to examine their planet radii and orbital eccentricities. However, this is difficult as there are currently only a handful of confirmed transiting warm Jupiters with measured orbits and masses.

We have initiated a ground-based follow-up campaign of warm Jupiter transiting candidates in order to increase the number of known transiting warm Jupiters that are confirmed as planets, i.e., their mass and orbit are measured. This campaign is part of a Las Cumbres Observatory (LCO; \citealt{brown13}) Key Project\footnote{ \href{http://web.gps.caltech.edu/\%7Eshporer/LCOKP/}{http://web.gps.caltech.edu/$\sim$shporer/LCOKP/}} (PI: Avi Shporer). Our current primary source of transiting candidates is the \kt\ mission \citep{howell14} and in the future the TESS mission \citep{ricker15, sullivan15}. We report here the confirmation of two \kt\ warm Jupiters orbiting \taro\ and \tart, which are the first discoveries from this project. Basic information about the two targets is given in \tabr{targets}. 

We give a more detailed motivation for our search for transiting warm Jupiter exoplanets in \secr{wj}. In \secr{obs} we describe our observations of \taro\ and \tart, in \secr{res} we describe our data analysis and results, and we discuss our new discoveries in \secr{dis}.

\begin{deluxetable}{lcc}
\tablecaption{Basic targets information \label{tab:targets}}
\tablewidth{0pt}
\tablehead{
\colhead{Parameter} & \colhead{\taro} & \colhead{\tart} 
}
\startdata
EPIC      & 211418729   & 211442297 \\
RA      & 08:31:31.911	& 08:26:12.827 \\
Dec      & 11:55:20.15 & 12:16:54.97 \\
$g$ [mag] & 15.07       & 13.59 \\
\kp\ [mag] & 14.29       & 13.19 \\
$r$ [mag] & 14.24       & 13.19\\
$i$ [mag] & 13.95       & 13.02 \\
$J$ [mag] & 12.84       & 12.11 \\
$H$ [mag] & 12.39       & 11.76 \\
$K$ [mag] & 12.30       & 11.72 \\
\enddata
\end{deluxetable}

%-------------------------------------------------------------------------------
\section{Why warm Jupiters?}
\label{sec:wj}
%-------------------------------------------------------------------------------

Our scientific goals are to (1) investigate the inflated gas giants conundrum, (2) study the mystery of hot Jupiters orbital evolution, and (3) identify targets for extending exoplanet atmosphere and stellar obliquity studies beyond the hot Jupiters class.
 
Briefly, the suggested mechanisms responsible for inflating gas giants can be divided into three categories \citep{weiss13, lopez16}: 
({\bf I}) Inflation due to stellar irradiation transported from the planet's atmosphere to its interior \citep[e.g.,][]{ginzburg16, komacek17} through, e.g., Ohmic dissipation \citep[]{batygin10}, thermal tides \citep{arras10}, kinetic energy transport \citep{showman02}, mechanical greenhouse \citep{youdin10}, or advection of potential energy \citep{tremblin17}.
({\bf II}) Inflation due to tidal heating following orbital eccentricity dissipation \citep{bodenheimer01}. If the planet eccentricity is continuously being excited, for example through interaction with a third body, this inflation mechanism can be long lived. 
({\bf III}) Delayed contraction, due to, e.g., increased atmospheric opacities \citep{burrows07}. Unlike the first two categories, the third category affects all giant planets, not only those on short or eccentric orbits.

One clue to understanding inflated gas giants is the empirical correlation between planet radius \rpl\ and stellar irradiation $f$ \citep[referred to hereafter as the radius-irradiation correlation, or $\rpl-f$ correlation; e.g.,][]{fortney07, laughlin11, enoch12, lopez16}. While it is consistent with inflation through irradiation it does not identify which of the category I mechanisms listed above is the dominant one, and, we must keep in mind that correlation does not necessarily mean causation.
The fact that the vast majority of known transiting gas giant exoplanets are at short periods, typically within 10 days or with irradiation above $10^8$~\ergscm, hinders a detailed understanding of the $\rpl-f$ correlation. 
While it seems that the correlation levels off at some irradiation level, the exact behavior is not clear, raising questions such as: {\it How low in irradiation does this correlation stretch? Is the lack of inflated warm Jupiters a robust feature? What is the irradiation below which there are no inflated gas giants? What drives the scatter in the correlation?}

Regarding hot Jupiters orbital evolution (goal 2 above), several theories attempt to explain how gas giants reach short orbital periods. Some invoke interaction with another object in the system (another planet or a stellar binary companion) where the gas giant planet is first injected into an eccentric orbit which then undergoes tidal circularization \citep[e.g.,][]{rasio96, fabrycky07}. Other theories suggest processes where the gas giant planet moves to an inward orbit by interacting with the protoplanetary disk, during which the orbit is kept circular \citep[e.g.,][]{lin96, alibert05}. Therefore, the two types of scenarios above differ in the orbital eccentricity of the gas giant planet as it is migrating from a wide orbit to a short orbit. Meaning, the orbital eccentricity of warm Jupiters is a clue to identifying the dominant orbital evolution channel of hot Jupiters, and is another motivation for expanding the known warm Jupiters sample to support statistical analysis. Some evidence that a significant fraction of warm Jupiters have circular orbits was gathered by studying the occurrence of additional planets in systems containing warm Jupiters compared to systems containing hot Jupiters. The increased occurrence of short-period planet companions to warm Jupiters \citep[e.g.,][]{steffen12, dong14, huang16} suggests their orbits are circular since non-circular orbits are expected to make the multi planet system dynamically unstable.

Another gap we wish to bridge with this program is the very small number of warm Jupiters available for exoplanet atmosphere and stellar obliquity studies (goal 3 above). For the bright stars in our sample detections of warm Jupiters will enable the study of exoplanet atmospheres with lower equilibrium temperature, and the study of stellar obliquity in star-planet systems with weaker tidal interaction.

%===============================================================================
\section{Observations}
\label{sec:obs}
%===============================================================================

%-------------------------------------------------------------------------------
\subsection{\kt\ photometry}
\label{sec:k2}
%-------------------------------------------------------------------------------

The two targets were initially identified as transiting planet candidates in \kt\ Campaign~5 photometry. They were observed by \kt\ in long cadence (30 minutes integration time) from 2015 April 27 to 2015 July 10. We reduced the \kt\ light curves following \cite{vanderburg14} and \cite{vanderburg16}. We then looked for transit signals using the Box-Least-Squares periodogram search \citep{kovacs02}, as implemented by \cite{vanderburg16}. Upon identifying transit candidates, we checked that they do not show known signs of a false positive by looking at the centroid motion of the target star during transit, searching for secondary eclipses, inspecting each individual transit, and confirming that the transit signal does not change significantly in light curves extracted from different photometric apertures. We then re-processed the light curves by simultaneously fitting for the transits, \kt\ thruster systematics, and low-frequency variations using the method described by \cite{vanderburg16}. The complete, detrended and normalized \kt\ Campaign 5 light curves are shown in \figr{lcfull} and the phase folded transit light curves in \figr{lc}. No additional transit signals were identified in the \kt\ light curve of both targets, nor was a secondary eclipse signal detected. 

\begin{onecolumn}
\begin{figure}
\includegraphics[scale=0.47]{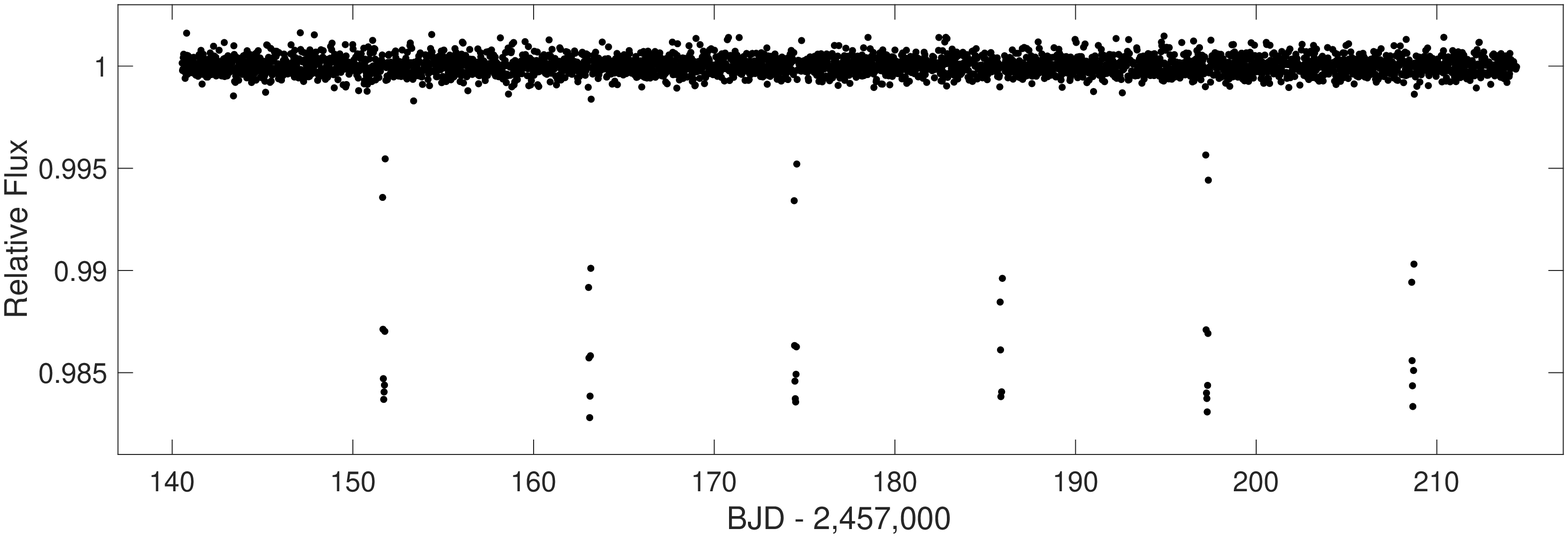}
\includegraphics[scale=0.47]{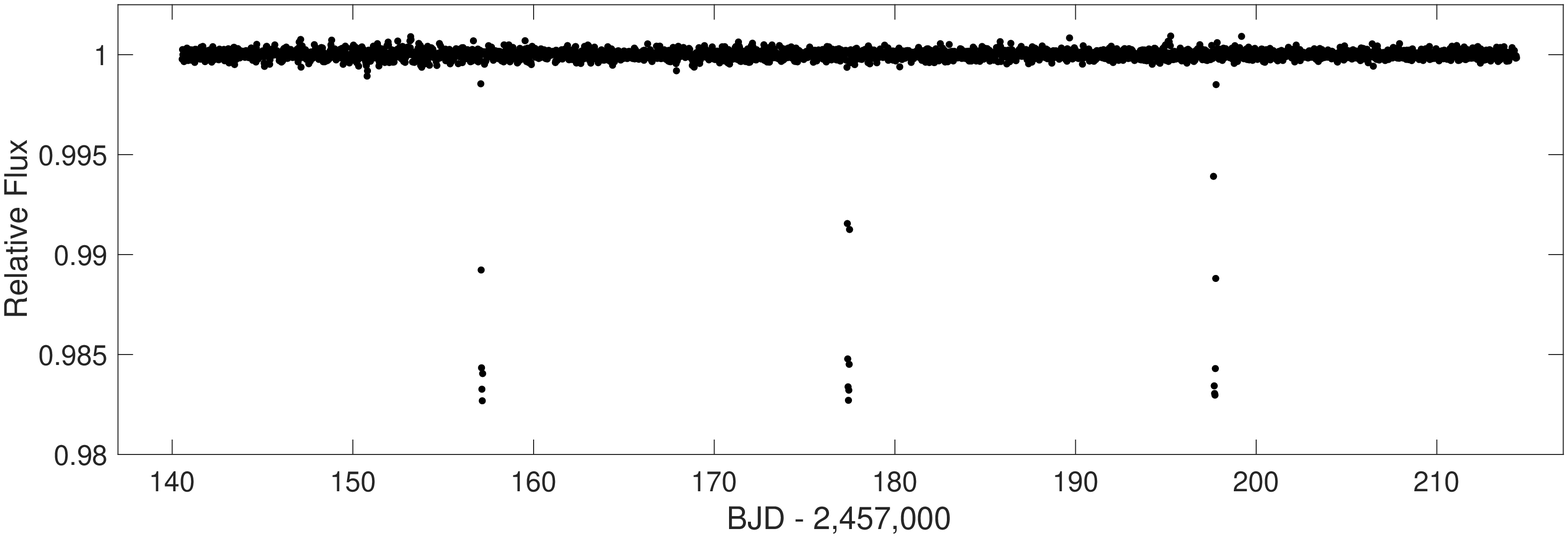}
\caption{Complete, detrended, and normalized \kt\ Campaign 5 light curves of \taro\ (top), including 6 transit events, and of \tart\ (bottom), including 3 transit events.
\label{fig:lcfull}}
\end{figure}
\end{onecolumn}

\begin{twocolumn}

We have searched for the host stars' rotation period by calculating their \kt\ light curves autocrrelation \citep{mcquillan14}. We used the non-detrended \kt\ light curves here while masking out the in-transit data. These light curves are shown in \figr{lcrot} along with the autocorrelation function. The periodicity detected for \taro\ is close to the \kt\ time span where it is challenging to separate stellar variability and long-term systematic features in the \kt\ data. Therefore we do not claim a detection of the rotation period for that star. For \tart\ we identify a 22.2 day periodicity, visually identified in the non-detrended light curve, interpreted as the host star's rotation period. We estimate this rotation period uncertainty to be at the 10 \% level to account for differential rotation \citep[e.g.,][]{reinhold13}, since during the limited \kt\ data time span the star spots are likely to be located within a narrow latitude range which is not known. \tart\ host star rotation period is longer but close to the 20.3 days orbital period. Although, given its mass and relatively long orbital period the planet is not expected to tidally synchronize the star's rotation \citep[e.g.,][]{mazeh08}.

\begin{figure}
\includegraphics[scale=0.4]{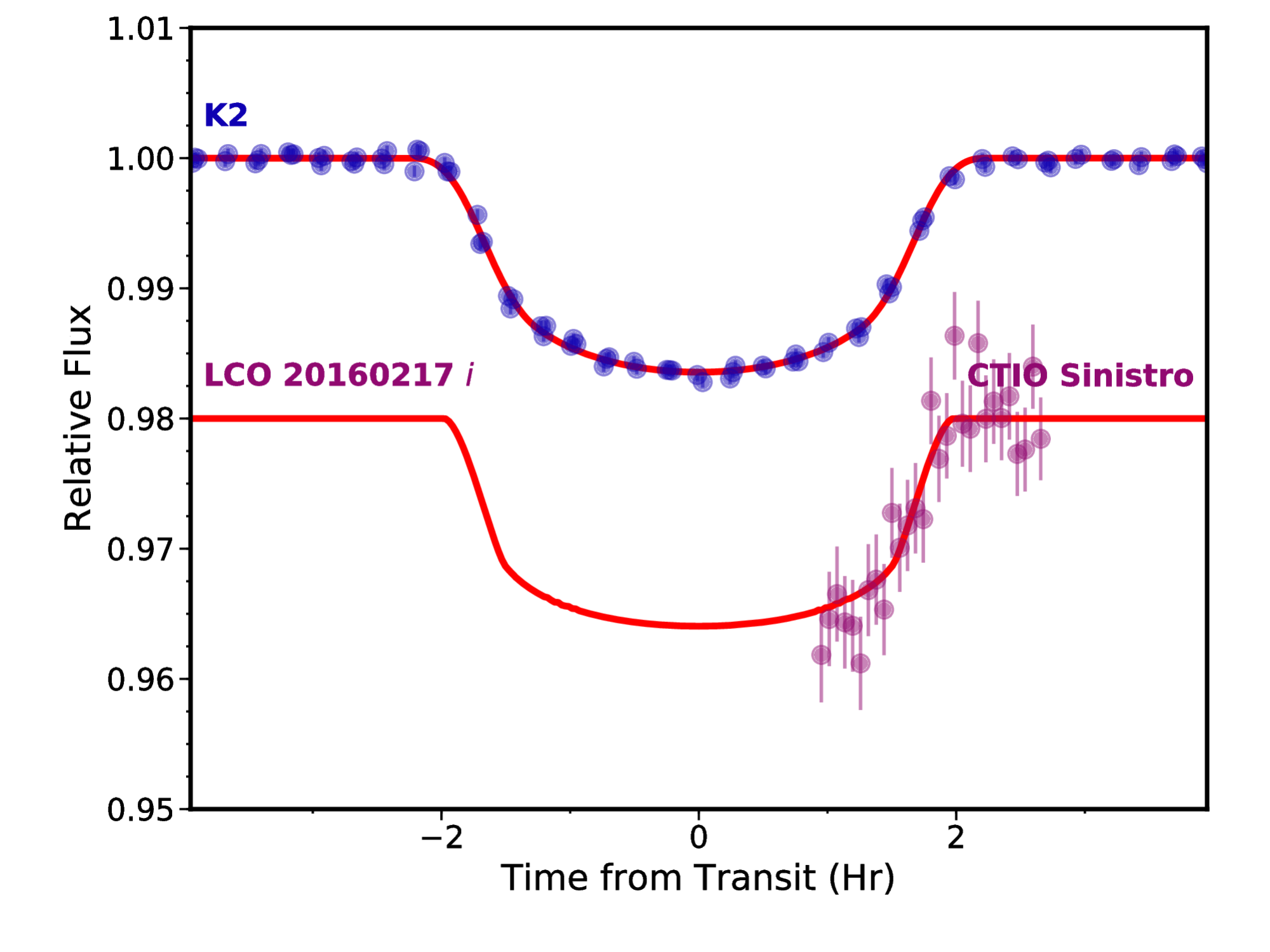}
\includegraphics[scale=0.4]{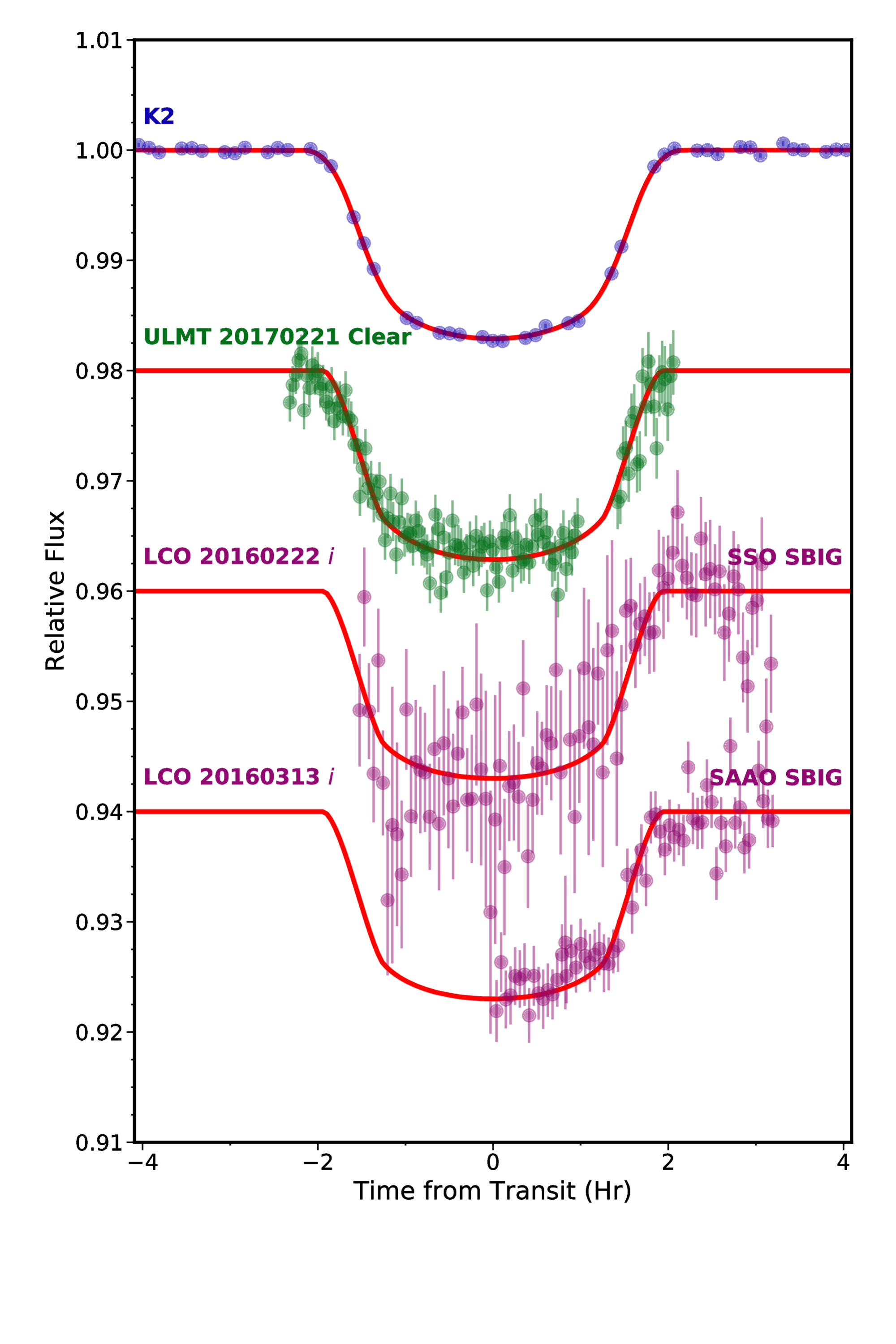}
\caption{Transit light curves of \tarob\ (top) and \tartb\ (bottom). The \kt\ light curves are phase folded and plotted in blue, ground-based follow-up light curves are plotted and labeled below and are arbitrarily offset in flux for visibility. The solid red line is the best-fit global model.
\label{fig:lc}}
\end{figure}

\end{twocolumn}
\begin{onecolumn}

\begin{figure}
\includegraphics[scale=0.51]{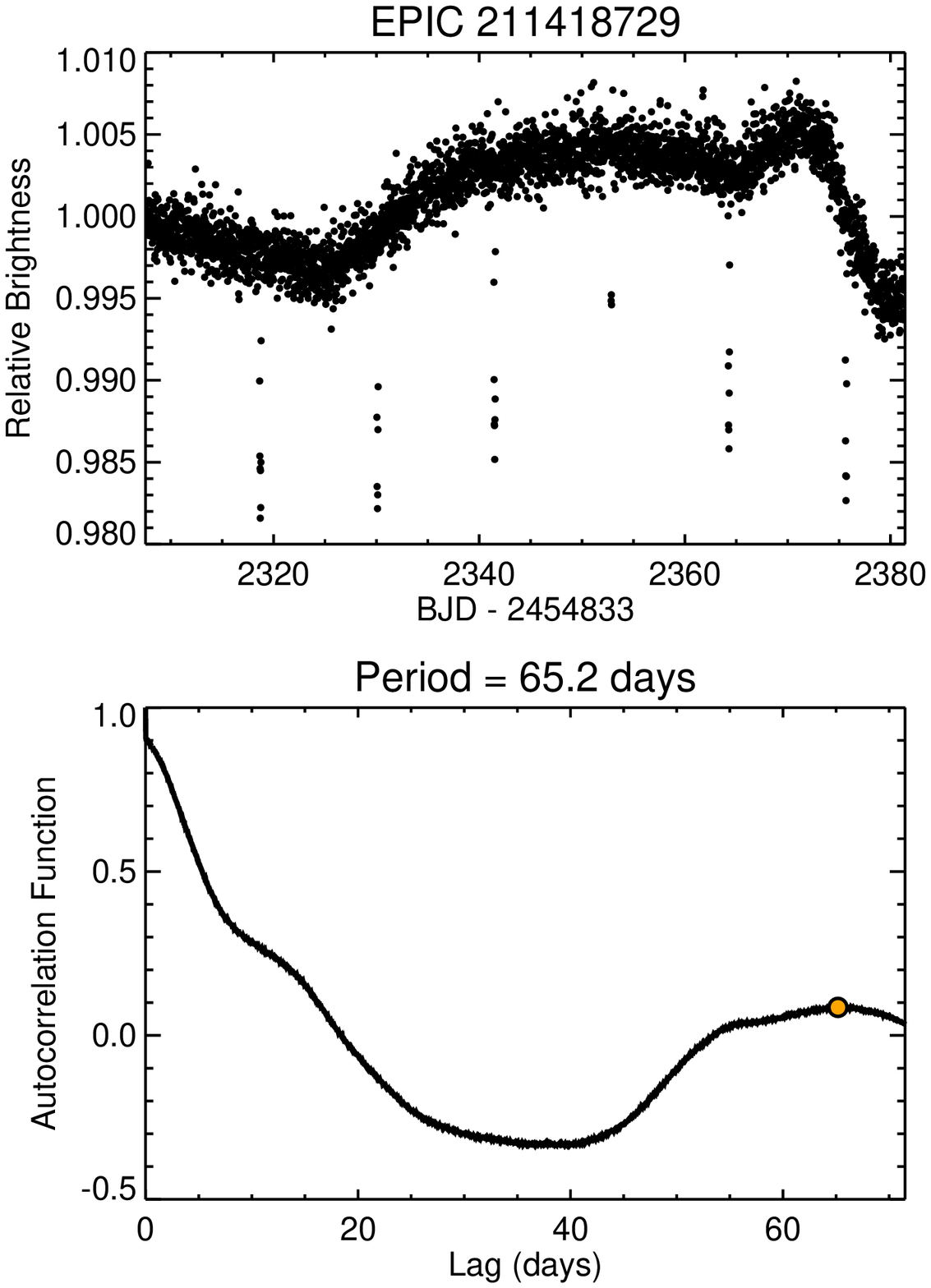}
\includegraphics[scale=0.51]{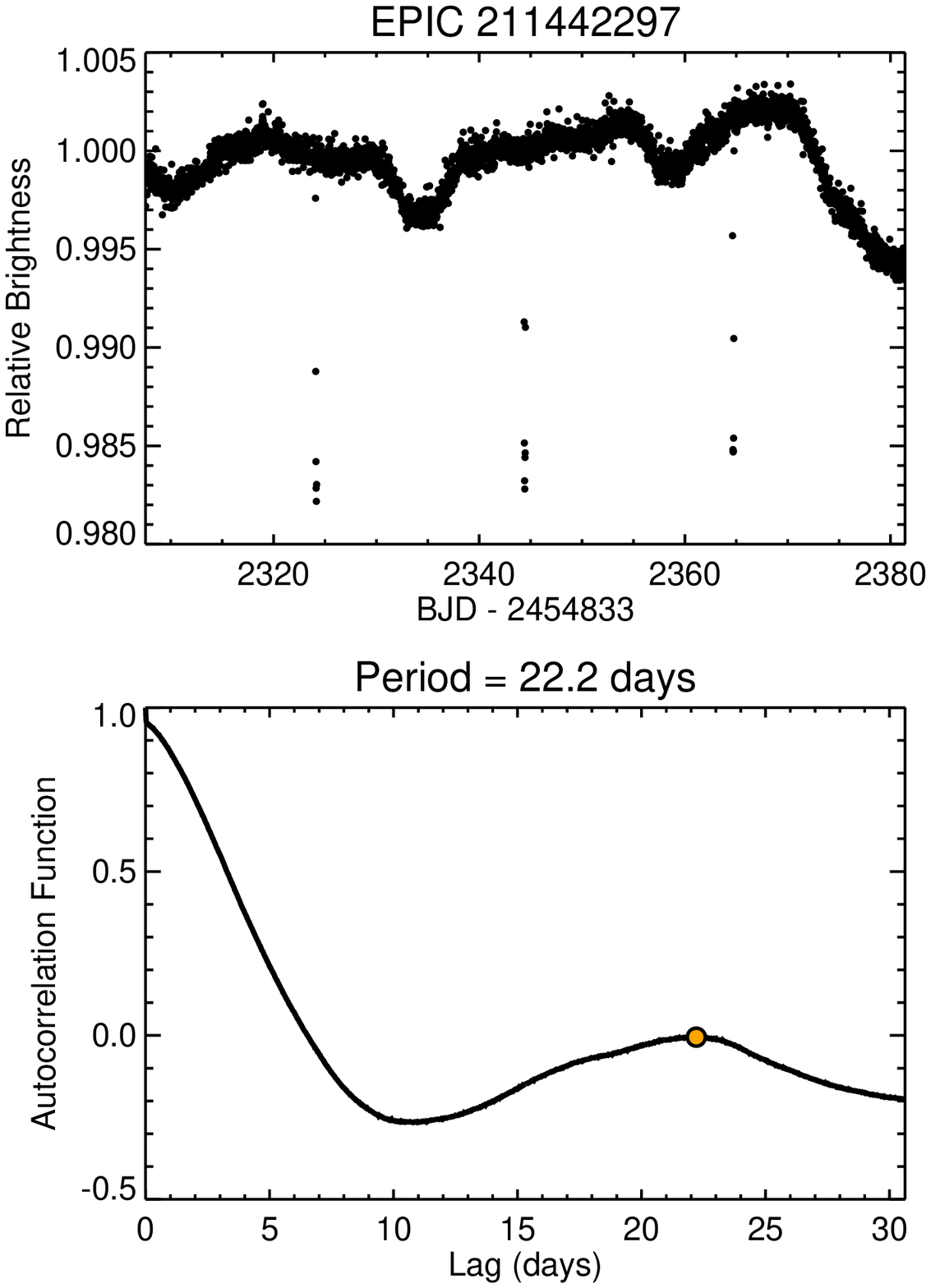}
\caption{\taro\ (left) and \tart\ (right) \kt\ light curves without detrending (top panels), and the light curve's autocorrelation (bottom panels) where the strongest peak is marked by an orange circle, and the period is listed in the panels' title. The in-transit photometric data was removed while calculating the autocorrelation function. For \taro\ the 65.2 day periodicity identified through the autocorrelation is close to the \kt\ data time span. At such time scales it is difficult to separate between stellar variability and systematic features in \kt\ data, hence we claim a detection of the rotation period only for \tart\ where the 22.2 day variability is identified visually in the light curve. We estimate a 10 \% uncertainty on that rotation period due to differential rotation.
\label{fig:lcrot}}
\end{figure}

\end{onecolumn}
\begin{twocolumn}

%-------------------------------------------------------------------------------
\subsection{High angular resolution imaging}
\label{sec:imaging}
%-------------------------------------------------------------------------------

After identifying the transit candidates in \kt\ photometry we checked that they are isolated targets using high angular resolution imaging. The imaging data was acquired as part of wider programs using Keck II and Gemini-North to obtain infrared adaptive optics (AO) and optical speckle imaging. The AO observations utilized the target stars as natural guide stars; the NIRC2 camera was utilized on Keck II and the NIRI camera was utilized on Gemini-North. NIRC2 has a pixel scale of 0.00942"/pixel and NIRI, with the Altair AO system, has a pixel scale of 0.0214"/pixel \citep{hodapp03}. The Gemini-North speckle observations were obtained with the visiting instrument DSSI which has a pixel scale of 0.011"/pixel \citep{horch09, horch12}.

For \taro\ the Keck AO data were obtained on 2016 February 19 with the Kp filter and an integration time of 17~s per frame for a total of 153~s, and the Gemini-North AO data were obtained on 2016 February 20 with the K filter and an integration time of 5~s per frame for a total of 50~s. For \tart\ the Keck AO data were obtained on 2016 January 21 with the Br gamma filter and an integration time of 30~s per frame for a total of 270~s. The Keck AO data have resolutions of 0.06--0.07" (FWHM) and the Gemini AO data have a resolution of 0.09" (FWHM).

The speckle data were obtained only for \tart, with the 692~nm and 883~nm filters on 2016 January 13. The data were obtained with 1000 60~ms frames in both filters  simultaneously. The speckle imaging is produced through a Fourier reconstruction of the speckle interferogram and have an angular resolution of 0.02" \citep{horch12}.

The sensitivity of the AO data was determined by injecting fake sources into the final combined images with separations from the primary targets in integer multiples of the central source's FWHM \citep{furlan17}. For the speckle data, the sensitivity was estimated from the scatter in the reconstructed image \citep{horch11, furlan17}. In both cases the sensitivity curves (contrast curves) represent \sig{5} limits and are shown in \figr{imaging}.  

For both targets, no stellar companions were detected in either the infrared AO or the optical speckle imaging, indicating (to the limits of the data) that the stars appear to be single stars with no additional components to either dilute the transit depths or confuse the determination of the origin of the transit signal \citep[e.g.,][]{ciardi15}.

%-------------------------------------------------------------------------------

\begin{figure}
\includegraphics[scale=0.4]{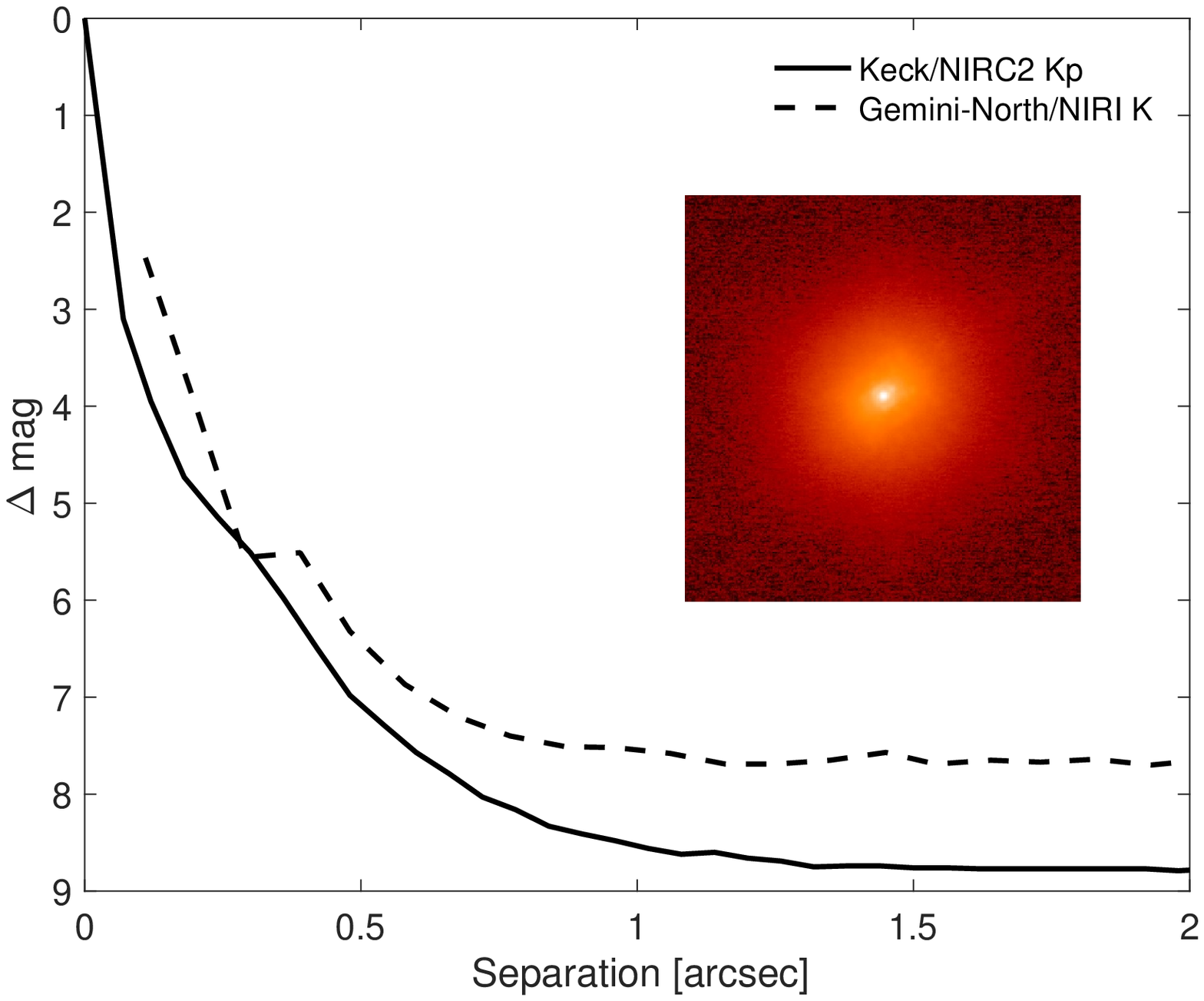}
\includegraphics[scale=0.4]{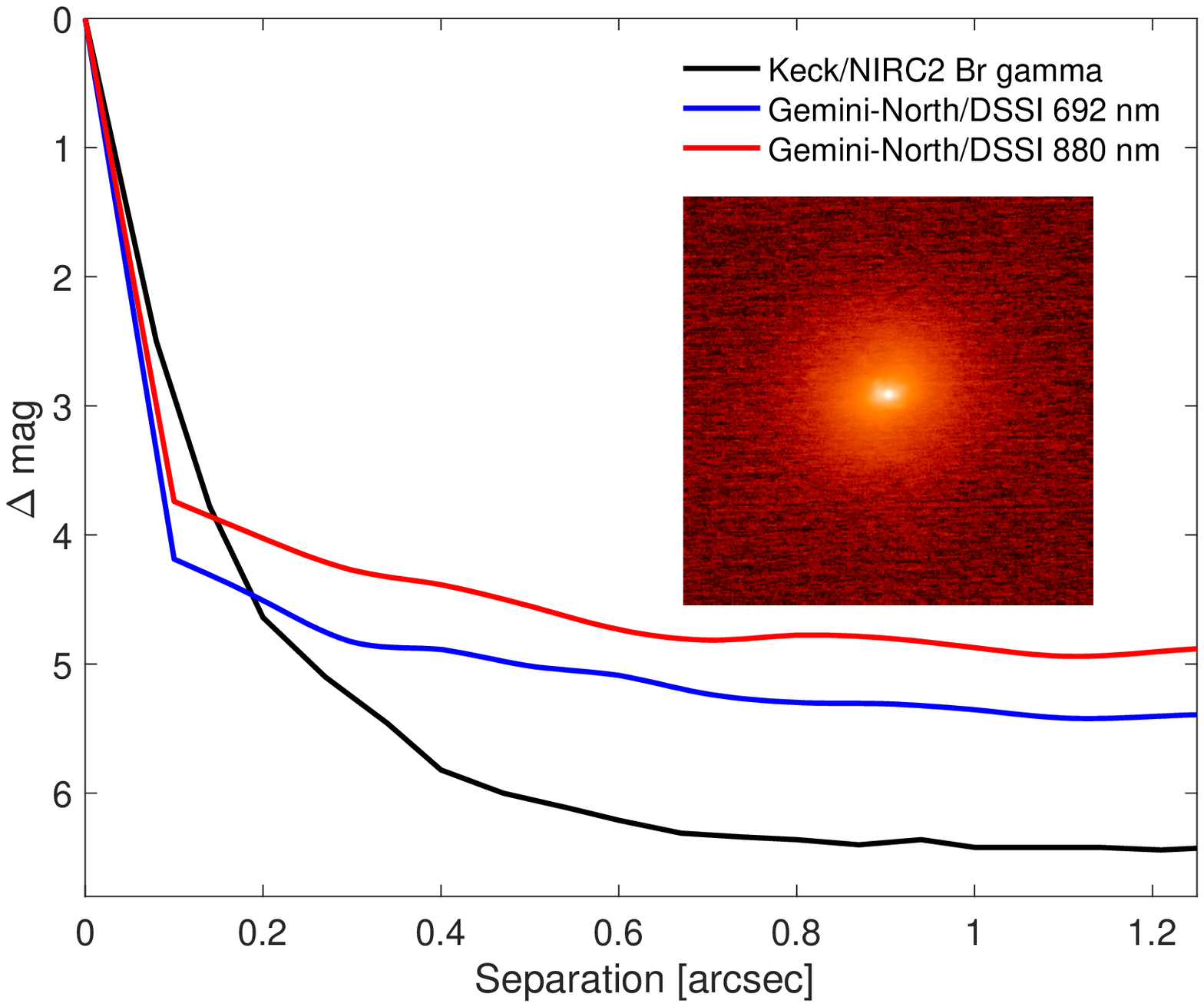}
\caption{Top: \taro\ contrast curves obtained by Keck/NIRC2 (black solid line; $Kp$ filter) and Gemini-North/NIRI (dashed black line; $K$ filter). Bottom: \tart\ contrast curves, obtained by Keck/NIRC2 in the Br gamma filter (black solid line) and Gemini North with DSSI (speckle imaging) in the 692 nm (blue solid line) and 880 nm (red solid line). In both panels the insets show the image obtained by Keck/NIRC2 in $Kp$ band spanning 3" on the side. 
\label{fig:imaging}}
\end{figure}

%-------------------------------------------------------------------------------
\subsection{Keck/HIRES spectroscopy and RV monitoring}
\label{sec:keck}
%-------------------------------------------------------------------------------

Once we confirmed that the two targets appear isolated in high angular resolution imaging we initiated gathering high resolution spectra with the HIRES instrument mounted on the Keck I telescope \citep{vogt94}. 

We collected a total of 6 spectra of \taro\ between 2016 February 2 and 2017 May 13. The first measurement was collected without the iodine cell in the light path for spectral characterization and searching for secondary lines \citep{kolbl15}. The remaining 5 spectra were observed with the iodine cell in the light path which imprints a dense forest of molecular absorption lines to be used as a simultaneous wavelength and instrumental point spread function reference, in order to measure the target's radial velocity (RV). The SNR for each spectrum was $\approx$40 per pixel and exposure times were typically 20 min. All spectra were collected using the 0.86''$\times$14'' slit for a resolution of about 65,000.

We collected 8 spectra and 7 RV measurements using the iodine cell for \tart\ between 2016 February 4 and 2017 April 10. Our setup was identical to that for \taro\ except that exposure times were generally shorter, $\approx$10 min, and SNR was slightly higher, $\approx$45 per pixel.

RVs for both stars were extracted by forward-modeling the composite iodine+stellar spectra in $\approx$800 small spectral chunks following the method of \cite{butler96}. We used the \texttt{SpecMatch} package \citep{petigura15} to synthesize an iodine-free stellar spectrum to be used in the modeling process \citep{fulton15} instead of collecting an expensive, long, high SNR iodine-free exposure of the target. 

We also used \texttt{SpecMatch} to extract the spectroscopic stellar parameters from our single iodine-free observation of each star. Those include the effective temperature \teff, surface gravity \logg, metallicity \feh, and stellar rotation projected on the line-of-sight \vsini\ where $V$ is the equatorial rotation and $I$ is the stellar rotation inclination angle. The \texttt{SpecMatch} results for the spectral parameters of both targets are listed in \tabr{spec}, and the RVs of both targets are listed in \tabr{rv} and plotted in \figr{rvcurves}.

In addition, we have calculated the activity indicators $S_{HK}$ and $\log R^{\prime}_{HK}$ \citep{isaacson10} for each of the Keck/HIRES spectra, and list in \tabr{spec} the mean of these indicators for the two host stars. For the indicators' uncertainties we adopt the scatter (standard deviation) in each sample. That scatter is affected primarily by the low SNR of the spectra in the Ca H \& K lines, which is 3--8 for \taro\ and 8--18 for \tart. Although \taro\ activity indicators suggest it is more active than \tart, we have detected a rotation period for the latter and not for the former (See \secr{k2}). This could be because the rotation period is too long to be detected in \kt\ data, consistent with the star being a slow rotator (\vsini\ $<$ 2 \kms). As seen in \figr{lcrot} we have identified a possible periodicity for \taro\ of 65.2 days, but, the proximity to the \kt\ time span means that we cannot reliably determine that that variability is stellar and not related to long-term systematic features.

%-------------------------------------------------------------------------------

\begin{deluxetable}{lcc}
\tablecaption{Spectroscopic parameters \label{tab:spec}}
\tablewidth{0pt}
\tablehead{
\colhead{Parameter} & \colhead{\taro} & \colhead{\tart} 
}
\startdata
\teff\ [K]     & 5014 $\pm$ 60   & 5544 $\pm$ 60    \\
\logg\ [cgs]   & 4.42 $\pm$ 0.07 & 4.33 $\pm$ 0.07  \\
\feh\  [dex]   & 0.41 $\pm$ 0.04 & $-$0.23 $\pm$ 0.04 \\
\vsini\ [\kms] & $<$ 2           & $<$ 2            \\
$S_{HK}$   & 0.258 $\pm$ 0.051 & 0.172 $\pm$ 0.014 \\
$\log R^{\prime}_{HK}$ [dex] & $-$4.854 $\pm$ 0.13 & $-$4.966 $\pm$ 0.078 \\
\enddata
\end{deluxetable}

\begin{deluxetable}{crc}
\tablecaption{Keck/HIRES radial velocities \label{tab:rv}}
\tablewidth{0pt}
\tablehead{
\colhead{Time} & \colhead{RV} & \colhead{RV err} \\
\colhead{BJD} & \colhead{\ms} & \colhead{\ms} 
}
\startdata
\multicolumn{3}{l}{\taro}\\
2457422.89476 & 152.4 &   5.8\\
2457789.92031 & -50.5 &   6.1\\
2457802.86068 &-194.3 &   6.8\\ 
2457853.77744 & 104.4 &   5.6\\
2457886.80575 & -32.3 &   8.2\\
\tableline\\
\multicolumn{3}{l}{\tart}\\
2457422.86801 &  -8.9 &   6.5\\
2457774.91483 &  37.4 &   7.3\\
2457789.90837 & -74.0 &   7.1\\
2457802.87129 &  91.4 &   7.1\\
2457804.81710 &  35.1 &   7.3\\
2457830.82364 & -58.9 &   7.2\\
2457853.76652 & -33.7 &   7.8\\
\enddata
\end{deluxetable}

\begin{figure}
\includegraphics[scale=0.4]{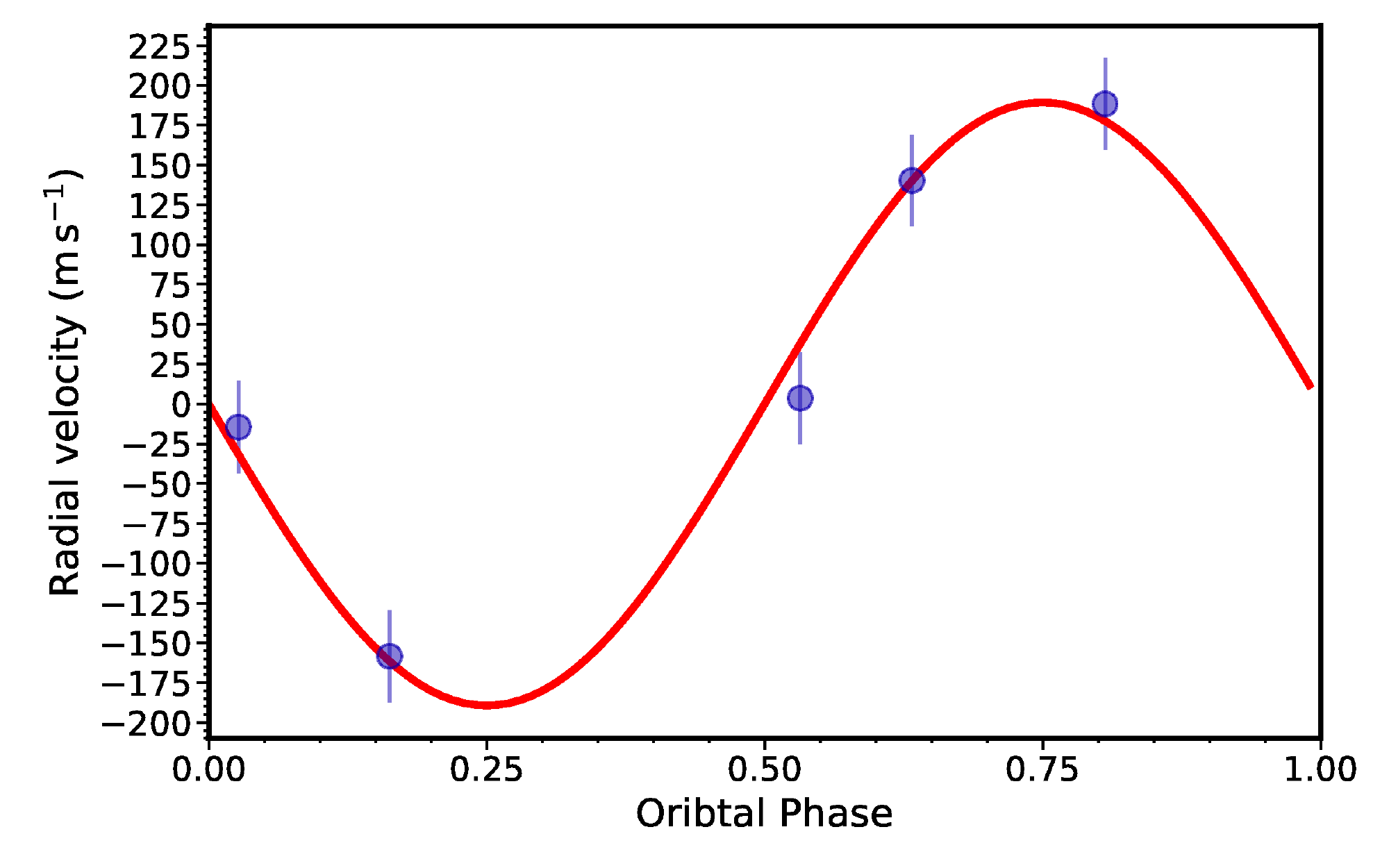}
\includegraphics[scale=0.4]{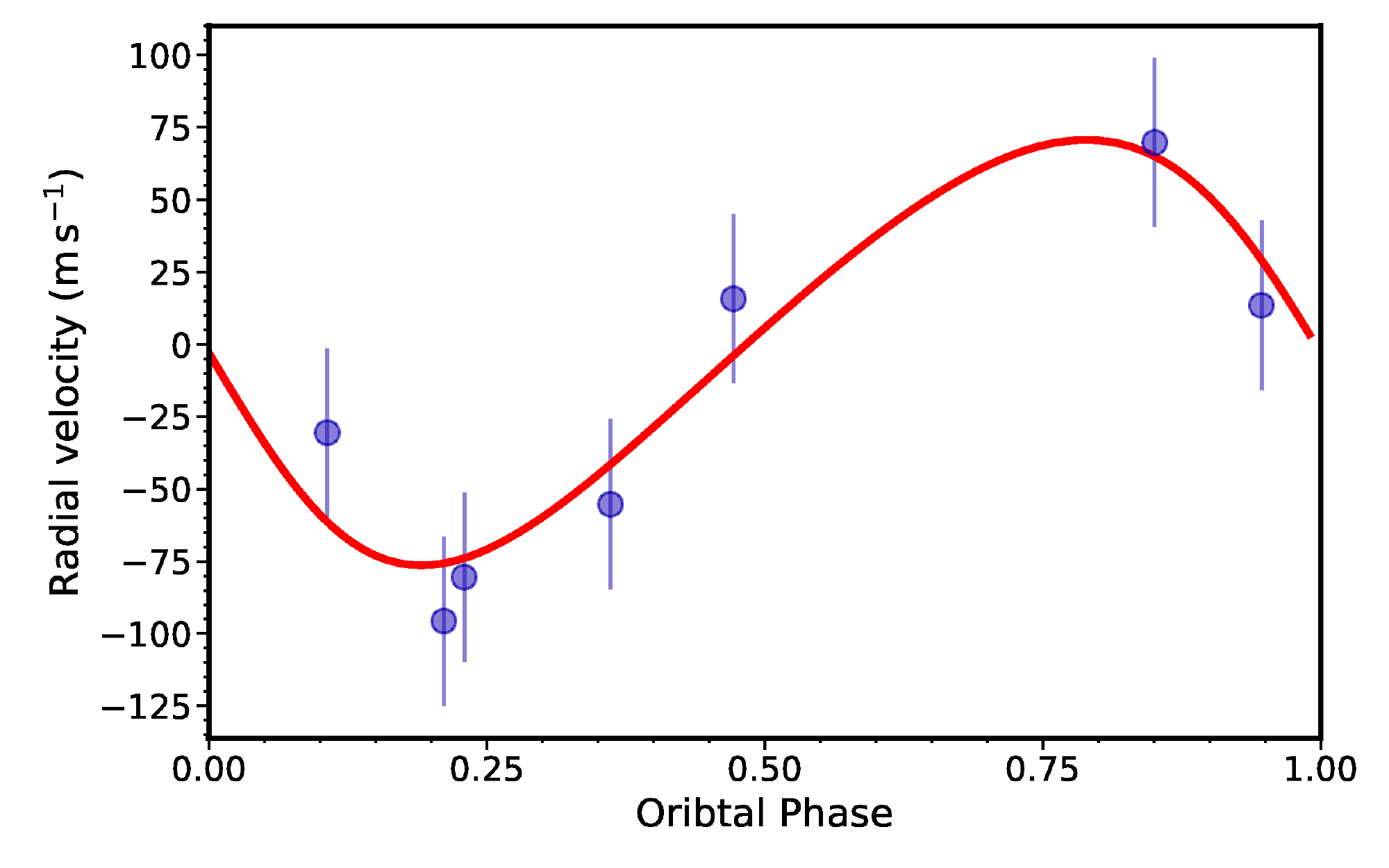}
\caption{Phase folded Keck/HIRES RV curves of \taro\ (top) and \tart\ (bottom). RVs are marked in blue circles and error bars and the fitted model is marked in a red solid line. The error bars shown in the plots include the jitter term added in quadrature (see \secr{res}).
\label{fig:rvcurves}}
\end{figure}

%-------------------------------------------------------------------------------
\subsection{Ground-based photometry}
\label{sec:phot}
%-------------------------------------------------------------------------------

While gathering RVs of the two targets we also acquired ground-based light curves of additional transit events. The target's brightness and transit depth seen in \kt\ data makes these transits observable from the ground using 1~m-class telescopes. Additional ground-based transits improve both the precision and accuracy of the transit ephemeris. The precision is improved due to the long time span between the transits observed by \kt\ and those observed from the ground, which is at least several times longer than a \kt\ campaign. The transit ephemeris accuracy is improved using additional ground-based transits since the half an hour sampling of the \kt\ long cadence data can lead to a biased ephemeris in case of an outlier measurement during one of the transits ingress or egress \citep{benneke17}. This is caused by a combination of the small number of transits within a \kt\ campaign for objects with relatively long periods as studied here, and the duration of the ingress/egress being comparable to the \kt\ long cadence integration time (30 minutes). 

These observations were done with the LCO network of 1~m telescopes and the KELT follow-up network, as described below.

%-------------------------------------------------------------------------------
\subsubsection{Las Cumbres Observatory (LCO)}
\label{sec:lco}
%-------------------------------------------------------------------------------

LCO is a fully robotic network of telescopes deployed in 6 sites around the globe in both hemispheres \citep[2 more sites are planned to be added by 2018;][]{brown13}. 

The egress of \tarob\ was observed on 2016 February 17 using a 1~m telescope at Cerro Tololo Inter-American Observatory (CTIO), Chile. The camera used was the Sinistro custom built imaging camera, with back-illuminated 4K$\times$4K Fairchild Imaging CCD with 15~\mic\ pixels (CCD486 BI). With a plate scale of 0.387"/pixel, the Sinistro cameras deliver a field of view (FOV) of 26.6'$\times$26.6'. The cameras are read out by four amplifiers with a readout time of $\approx$45~s. We used the $i$-band filter, with exposure time of 180~s and a slight defocus of the telescope (1\,mm) to spread out the PSF over more pixels and eliminate the risk of saturation at the core of the PSF. Images were reduced by the standard LCO pipeline \citep{brown13}, and aperture photometry was performed in the manner set out in \cite{penev13} through a fully automated pipeline developed in our group (Espinoza et al.~2017, in prep.). 

For \tartb\ LCO observed an almost complete transit on 2016 February 22 with a 1~m telescope at Siding Spring Observatory (SSO), Australia, and an egress on 2016 March 13 with a 1~m telescope at South African Astronomical Observatory (SAAO), South Africa. These two observations were made using the older SBIG cameras (which have since been replaced by Sinistro cameras). The SBIG cameras featured front-illuminated 4K$\times$4K KAF-16803 CCDs with 9~\mic\ pixels. With a plate scale of 0.232"/pixel these cameras have a FOV of 15.8'$\times$15.8'. We used 2$\times$2 pixel binning which results in a readout time of 15.5 s. We again used the $i$-band, exposure times of 180\,s and a telescope defocus of 1\,mm. Data was reduced to light curves in the same manner as set out for the Sinistro camera reduction.  We note that the combination of a smaller FOV (fewer reference stars) and front-illuminated CCDs (large intra-pixel variation and lower quantum efficiency) means that the precision of the light curves derived from the SBIG cameras is much lower than the Sinistro cameras. Additionally, site conditions at CTIO are typically much better for photometry than at either SAAO or SSO, which also contributes to the precision of the photometry.

The LCO light curves are shown in \figr{lc} and listed in \tabr{gbphot}.

%-------------------------------------------------------------------------------
\subsubsection{University of Louisville Manner 0.6~m Telescope}
\label{sec:ulmt}
%-------------------------------------------------------------------------------

We observed one full transit of \tartb\ with short segments of out-of-transit baseline using the University of Louisville Manner Telescope (ULMT) located at the Mt.~Lemmon summit of Steward Observatory, AZ, on 2017 February 21 with no filter. ULMT is a member of the KELT follow-up network, composed of 1~m-class telescopes and smaller telescopes dedicated to the photometric follow-up of transiting planet candidates. The observations employed a 0.6~m f/8 RC Optical Systems Ritchey-Chr\'{e}tien telescope and SBIG STX-16803 CCD with a 4k $\times$ 4k array of 9 $\mu$m pixels, yielding a $26\farcm6 \times 26\farcm6$ field of view and $0\farcs39$ pixel$^{-1}$ image scale. The telescope was defocused, resulting in a ``donut''-shaped stellar PSF and guiding was applied to maintain stable pointing.

The telescope control computer malfunctioned at about the time egress started, but was recovered about 25 minutes later. 

The image sequence was bias, dark, and flat-field corrected using AstroImageJ (AIJ; \citealt{collins17}). We also used AIJ to extract differential photometry using aperture photometry. 
An iterative $2\sigma$ cleaning routine was employed to exclude outlier pixels and pixels containing flux from nearby stars from the background region. To normalize the target light curve and have the out-of-transit flux level set to unity we used a comparison ensemble of ten stars that produced the lowest model fit residuals. 

The ULMT 0.6~m \tart\ light curve is shown in \figr{lc} and listed in \tabr{gbphot}.

%-------------------------------------------------------------------------------

\begin{deluxetable}{ccc}
\tablecaption{Ground-base photometry\label{tab:gbphot}}
\tablewidth{0pt}
\tablehead{
\colhead{Time} & \colhead{Relative} & \colhead{Relative} \\
\colhead{BJD} & \colhead{Flux} & \colhead{Flux error} 
}
\startdata
\multicolumn{3}{l}{\taro\ - LCO}\\
2457436.53253 &	0.9818 & 0.0036\\
2457436.53508 &	0.9846 & 0.0035\\
2457436.53763 &	0.9865 & 0.0036\\
\multicolumn{3}{l}{\tart\ - LCO}\\
2457440.91603 &	0.9910 & 0.0056\\
2457440.91825 &	1.0012 & 0.0050\\
2457440.92048 &	0.9909 & 0.0048\\
\multicolumn{3}{l}{\tart\ - ULMT 0.6~m}\\
2457805.79750 & 0.9979 & 0.0015\\
2457805.79880 & 0.9995 & 0.0015\\
2457805.80032 & 1.0005 & 0.0015
\enddata
\tablecomments{\tabr{gbphot} is published in its entirety in the electronic edition of the paper. A portion is shown here for guidance regarding its form and content.}
\end{deluxetable}

%===============================================================================
\section{Data analysis and results}
\label{sec:res}
%===============================================================================

To derive accurate parameters for each system we performed a global modeling of the available photometric and spectroscopic observations. We used the model fitting procedure described in \cite{zhou17}. This includes making use of the \kt\ photometry, ground-based follow-up light curves, RV measurements, and spectroscopic atmospheric properties of the host stars. The light curves are modeled as per \citet{mandel02}, where the free parameters are the orbital period $P$, mid transit time $T_0$, the planet to star radii ratio $\rpl/\rstar$, normalized orbital semi-major axis distance $a/\rstar$, line-of-sight orbital inclination $i$, and orbital eccentricity parameters $\sqrt{e}\cos \omega$ and $\sqrt{e} \sin \omega$ where $e$ is the eccentricity and $\omega$ the argument of periastron. The 30 minutes duration of the \kt\ long cadence exposures is accounted for by integrating over 10 model steps per exposure. Quadratic limb darkening coefficients are interpolated from \citet{claret04} to the atmospheric parameters of each star, and held fixed during the fitting. For the ULMT observation of \tart\ (\secr{ulmt}), obtained without a filter, we adopted the same limb darkening parameters as that of the \kt\ light curves. 

The RVs are modeled by a Keplerian orbit, with additional free parameters for the orbit RV semi-amplitude $K$ and systemic center-of-mass RV $\gamma$. We include a fitted jitter term $s$ to model the RVs as per \citet{haywood16}. Since the transit duration is dictated by the stellar density, we also make use of the precise \kt\ photometry to refine the stellar parameters. We interpolate the Dartmouth stellar isochrones \citep{dotter08} over the axes of stellar atmospheric effective temperature $\teff$, mean stellar density $\rho_s$, and metallicity \feh, to derive a surface gravity $\logg$. 

The transit-derived stellar density and spectroscopically constrained effective temperatures are plotted in \figr{rhostar} against Solar metallicity isochrones to illustrate this isochrone interpolation process. We reject solutions that yield system ages older than 13 Gyr, the age of the thin disk of the galaxy \citep{knox99}. At each iteration, we include a log-likelihood term calculated between the transit-derived $\logg$ with that measured from spectroscopy. The posterior probability distribution is explored via a Markov Chain Monte Carlo (MCMC) analysis, using the affine invariant sampler \emph{emcee} \citep{emcee13}. Gaussian priors are applied for the stellar atmospheric parameters $\teff$ and \feh, and all other parameters are assumed to follow uniform priors. 

For both objects we ran two fits. One that assumes the orbit is circular and the other that fits for the eccentricity $e$ and argument of periastron $\omega$. For \tarob\ the eccentric orbit fit did not give a statistically significant eccentricity, hence we adopt the circular orbit model. The upper limit on the orbital eccentricity are 0.06 and 0.41 at \sig{1} and \sig{3}, respectively. 

For \tart\ the eccentricity is measured at a statistical significance of close to \sig{2}, hence we adopted the eccentric orbit model. We tested the significance of the measured orbital eccentricity by refitting this system using a Beta function prior distribution on the eccentricity following \cite{kipping14}. That analysis gave consistent results at the \sig{1}\ level. Another reason for adopting the eccentric orbit model was that it gave consistent results for the derived stellar parameters (mass, radius, and age) with those derived when fitting using only stellar isochrones (with priors on the spectral parameters \teff, \logg, and \feh) and without the light curve and RV curve. The derived stellar parameters from the circular orbit model fit are not consistent with the results from fitting using stellar evolutionary models alone.

The 68\% confidence region for the model fit free parameters, as well as a series of inferred system parameters, are listed in \tabr{params}. The best fit transit light curve models are shown in \figr{lc} and the RV fits are shown in \figr{rvcurves}.

\begin{figure*}
\includegraphics[scale=0.64]{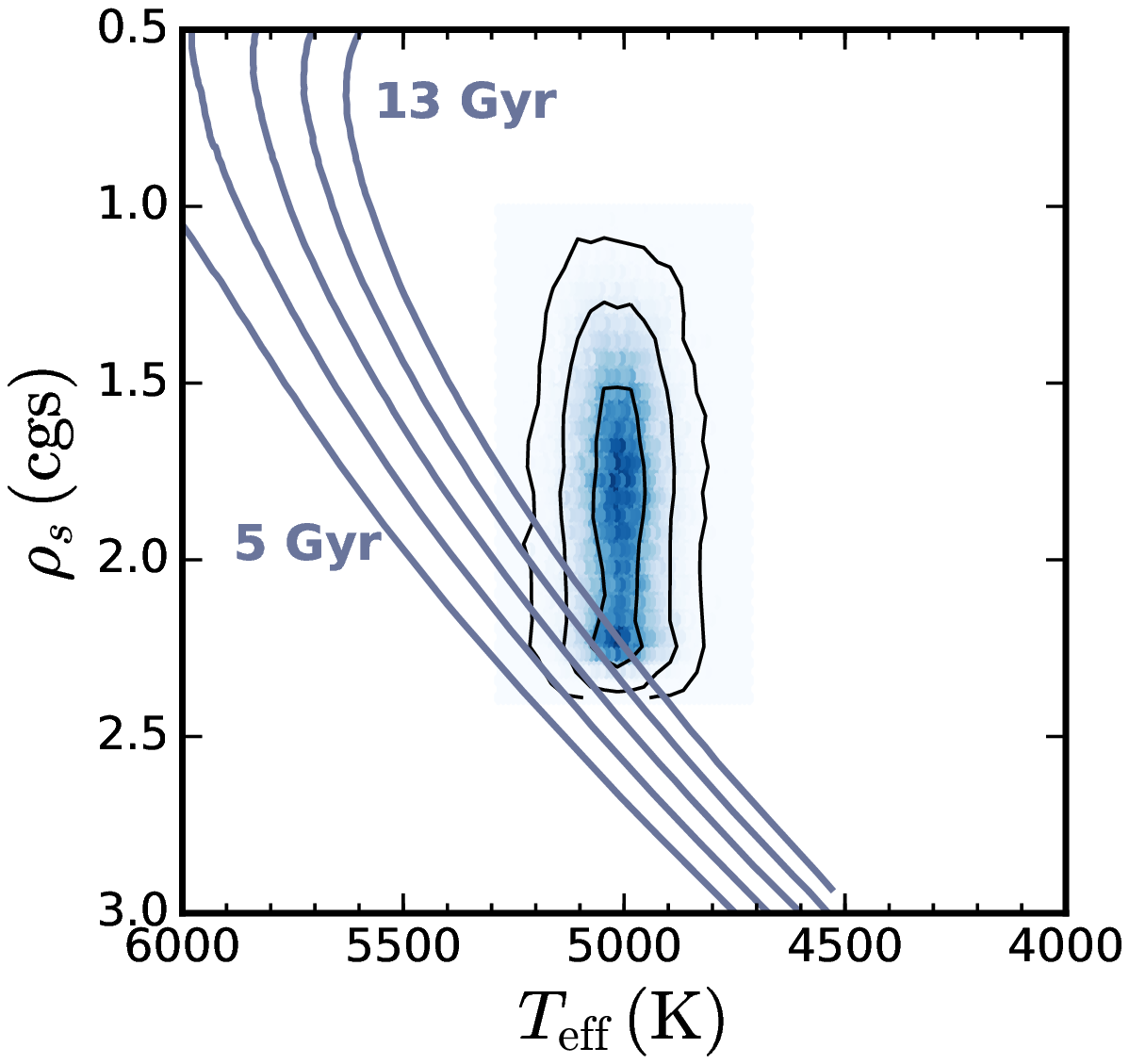}
\includegraphics[scale=0.64]{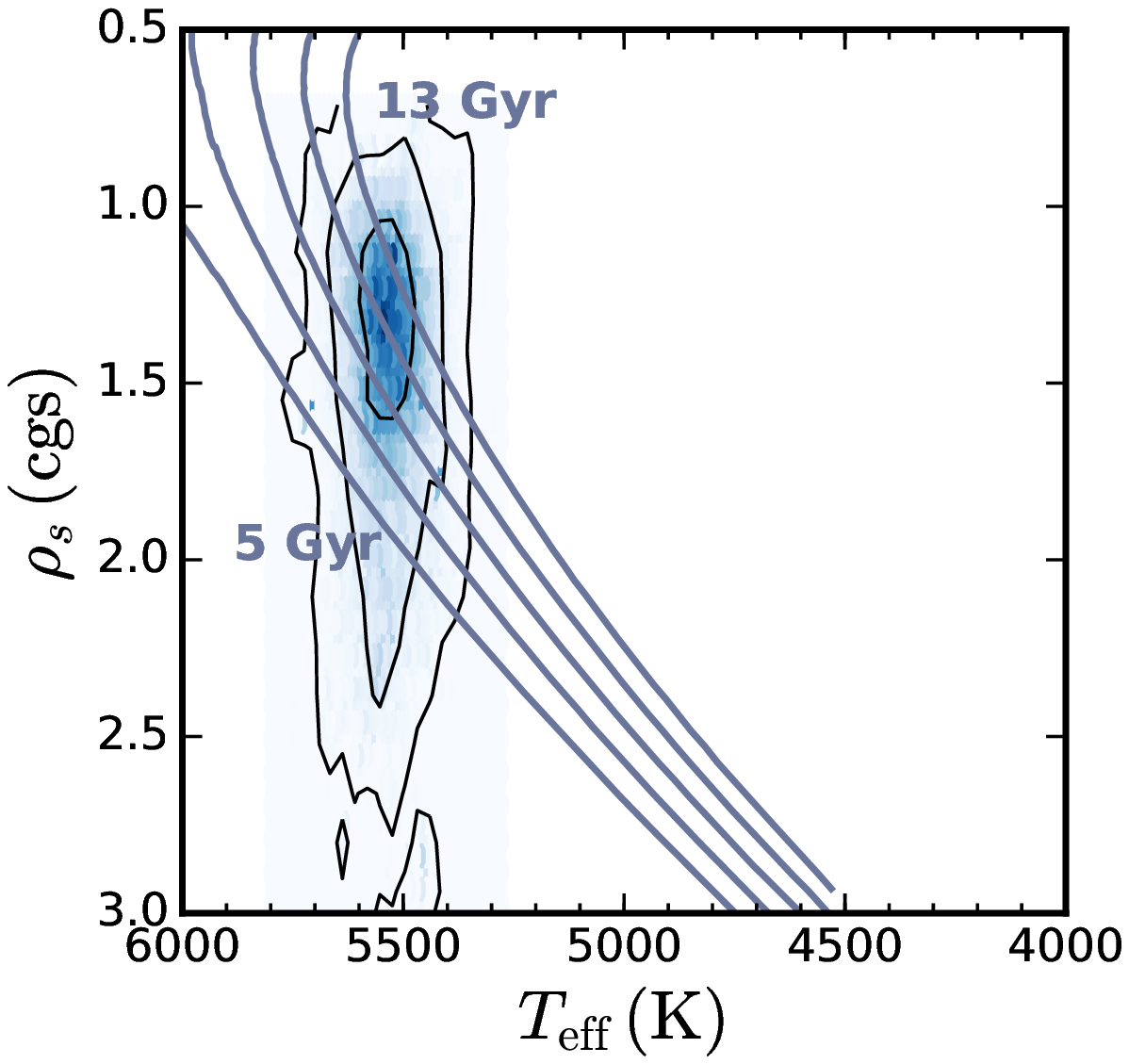}
\caption{Stellar parameters are derived via interpolation of the Dartmouth isochrones. The stellar density (\rhostar), derived from the transit light curve, and the spectroscopic effective temperature \teff\ and metallicity [Fe/H] are compared against isochrone tracks at each MCMC iteration to constrain the stellar properties. The stellar densities and effective temperatures of \taro\ (left) and \tart\ (right) are plotted. The Dartmouth solar metallicity isochrones at ages of 5, 7, 9, 11, and 13 Gyrs are drawn as a guide. Note that solutions yielding ages $>13$ Gyr are removed during the derivation of system and stellar parameters in the global analysis, as that age is older than the age of the thin disk of the galaxy.
\label{fig:rhostar}}
\end{figure*}

%-------------------------------------------------------------------------------
\begin{deluxetable}{lcccccc}
\tablecaption{Fitted and derived parameters \label{tab:params}}
\tablewidth{0pt}
\tablehead{
\colhead{Parameter} & \multicolumn{3}{c}{\taro} & \multicolumn{3}{c}{\tart}  \\ 
 & \colhead{Value} & \colhead{+\sig{1}} & \colhead{-\sig{1}} & 
   \colhead{Value} & \colhead{+\sig{1}} & \colhead{-\sig{1}} 
}
\startdata
\multicolumn{6}{l}{\it Fitted parameters}\\
$P$\ [day]       & 11.39109 & 0.00018 & 0.00017 & 20.273034 & 0.000036 & 0.000037 \\
$T_0$\ [BJD]     & 2457174.49729 & 0.00033 & 0.00033 & 2457157.15701 & 0.00025 & 0.00025 \\
$\gamma$\ [\ms]  & -36 & 14 & 14 & 22 & 15 & 12 \\
$K$\ [\ms]       & 189 & 21 & 22 & 77 & 17 & 16 \\ 
$\sqrt{e}\cos(\omega)$  & 0\,\tablenotemark{a} & - & - & -0.12 & 0.25 & 0.18 \\
$\sqrt{e}\sin(\omega)$  & 0\,\tablenotemark{a} & - & - & 0.28 & 0.11 & 0.17\\
Jitter $s$ [\ms] & 29 & 12 & 25 & 24 & 16 & 8 \\
$a/\rstar$      &  24.44 & 0.42 & 0.63 & 33.8 & 2.3 & 1.7 \\
$R_p/\rstar$    & 0.11432 & 0.00102 & 0.00073 & 0.1254 & 0.0011 & 0.0011\\
$i$\ [deg]      & 89.53   & 0.30  & 0.25 & 88.82 & 0.15 & 0.15\\
\teff\  [K]     & 5027 & 62 & 57 & 5560 & 56 & 58  \\
\feh            & +0.410 & 0.037 & 0.035 & -0.220 & 0.035 & 0.036 \\
$u_1$\,\kt\,\tablenotemark{a}		& 0.5815 & -- & -- & 0.4430 & -- & --\\
$u_2$\,\kt\,\tablenotemark{a}		& 0.1392 & -- & -- & 0.2312 & -- & --\\
$u_1$\,ULMT\,\tablenotemark{a}		& -- & -- & -- & 0.4430 & -- & -- \\
$u_2$\,ULMT\,\tablenotemark{a}		& -- & -- & -- & 0.2312 & -- & -- \\
$u_1$\,$i'$\,\tablenotemark{a}    	& 0.4225 & -- & -- & 0.3118 & -- & -- \\
$u_2$\,$i'$\,\tablenotemark{a}	    & 0.2472 & -- & -- & 0.3047 & -- & -- \\
\hline
\multicolumn{6}{l}{\it Derived parameters}\\
\mstar\ [\msun] & 0.832 & 0.021 & 0.018 & 0.831 & 0.023 & 0.019  \\ 
\rstar\ [\rsun] & 0.828 & 0.026 & 0.022 & 0.881 & 0.049 & 0.050  \\ 
 {\rhostar\ [cgs]} &  {1.84} &  {0.32} &  {0.29} &  {1.43} &  {0.44} &  {0.28} \\
\logg\ [cgs] & 4.481 & 0.044 & 0.051    & 4.461 & 0.057 & 0.041 \\ 
Age [Gyr]       & 9.9 & 2.3 & 3.2     & 10.7 & 1.7 & 4.2 \\  
\rpl\ [\rjup] & 0.942 & 0.032 & 0.020   & 1.115 & 0.057 & 0.061 \\ 
\mpl\ [\mjup] & 1.85 & 0.23 & 0.22      & 0.84  & 0.18  & 0.20 \\ 
 {$\rho_p$ [cgs]} &  {2.99} &  {0.46} &  {0.45} &  {0.82} &  {0.30} &  {0.24} \\
$a$\ [au] & 0.09309 & 0.00066 & 0.00059  & 0.1367 & 0.0012 & 0.0010 \\ 
\teq\ [K]\,\tablenotemark{b} & 719 & 15 & 11             & 682 & 22 & 24 \\ 
$b$      & 0.20 & 0.10 & 0.12         & 0.702 & 0.047 & 0.053 \\ 
$T_{12}$\ [d] & 0.01732 &0.00109 & 0.0068  & 0.0343 & 0.0050 & 0.0046  \\ 
$T_{14}$\ [d] & 0.1627 &0.0011 & 0.0010  & 0.1679 & 0.0027 & 0.0046 \\
$e$             & 0\,\tablenotemark{a} & -- & --    & 0.137   & 0.072  & 0.074  \\ 
$\omega$\ [deg] & -- & -- & -- & 104 & 41 & 52 \\ 
 {Distance [pc]} &  {481} &  {20} &  {15} &  {417} &  {26} &  {25} \\
 {$A_V$ [mag]} &  {0.109} &  {0.072} &  {0.072} &  {$<$0.12\,\tablenotemark{c}} & -- & -- \\
\enddata
\tablenotetext{a}{ Parameter was fixed during the model fitting process.}
\tablenotetext{b}{ Assuming zero albedo and no redistribution of heat.}
\tablenotetext{c}{ $3\sigma$ upper limit given for reddening.}
\end{deluxetable}

%-------------------------------------------------------------------------------

%===============================================================================
\section{Discussion and Conclusions}
\label{sec:dis}
%===============================================================================

Both \tarob\ and \tartb\ are among the longest period transiting gas giant planets with a measured mass. In fact, according to the NASA Exoplanet Archive \citep{akeson13} \tartb\ is currently\footnote{As of 2017 June 1st.} the longest period \kt\ transiting exoplanet with a well constrained mass (but see \citealt{bayliss17}). 

The number of RVs we have accumulated for each system is relatively small, with 5 for \taro\ and 7 for \tart. The relatively small number of RVs  results in a relatively poor constraint of the orbital eccentricity and RV semi-amplitude. The latter has an uncertainty of 12\% for \taro\ and close to 22\% for \tart, leading to similar uncertainties on the two planet masses.

It is interesting to note that both host stars are relatively old, with ages close to 10~Gyr, although with the typical large age uncertainties (see \tabr{params}). For \tartb, despite the host star old age the combination of the measured orbital eccentricity ($e$=0.137$^{+0.072}_{-0.074}$) and orbital separation ($a/\rstar = 33.8^{+2.3}_{-1.7}$) suggests that if the orbit is indeed eccentric that eccentricity is primordial, since it is not expected to be tidally circularized within the host star's lifetime \citep[e.g.,][]{mazeh08}.

\figr{radmassflux} top panel shows the planet mass-radius diagram for gas giant plants, with $\rpl > 0.6\ \rjup$, and with well measured planet radius and mass. Those include 273 planets listed on the NASA Exoplanet Archive with planet radius error smaller than 0.15 \rjup\ and planet mass error below 20\% of the planet mass itself. Not included in that sample are circumbinary planets and directly imaged planets. The black and gray solid lines show the range of theoretical planet radius where the planet radius grows as the mass of its rocky core decreases \citep{fortney07}. The dashed gray lines are equal mean density lines. The two new planets, \tarob\ and \tartb\ are marked in red. Both planets are not inflated compared to theoretical expectations, unlike many other planets in the diagram. Their positions are close to or consistent with theoretical expectations for a planet with little to no rocky core, for \tartb, and a planet with a significant rocky core for \tarob. 

The difference in the expected core mass between the two planets, combined with the larger planet mass of \tarob\ compared to \tartb, agrees with the empirical correlation between heavy element mass and planet mass for gas giants \citep{miller11, thorngren16}. The difference in the host stars metallicity, with \taro\ being super-solar and \tart\ being sub-solar, also agrees with the gas giant planet mass - host star metallicity correlation \citep{miller11, thorngren16}. These correlations allow to estimate the planets composition \citep{espinoza17}.

\figr{radmassflux} bottom panel shows the planet radius - stellar irradiation diagram (\rpl--$f$), including the same sample of planets as in the top panel, and where the two new planets are marked in red. Their positions are consistent with the hypothesis that the \rpl--$f$ correlation does not continue below irradiation of 10$^8$~\ergscm, where the correlation levels off and stellar irradiation does not significantly affect the planet radius. If true, this can be used as a clue for identifying the physical mechanism inflating gas giant planets, and, it makes warm Jupiters good targets for testing theoretical mass-radius relations as their radius is not affected by a physical mechanism that is currently not completely understood. However, an accurate characterization of the behavior of planet radius at low stellar irradiation requires the detection of many more warm Jupiters.

Finally, we note that the two new planets reported here are planned to be observed by \kt\ again during Campaign~18, from May to August 2018, when \kt\ will re-observe the Campaign~5 field\footnote{See list of \kt\ fields here: \href{https://keplerscience.arc.nasa.gov/k2-fields.html}{https://keplerscience.arc.nasa.gov/k2-fields.html}}. If successful, this will give a 3 year time span and therefore allow refining the transit ephemerides and the planet-to-star radii ratio, looking for transit timing variations and searching for other transiting planets in those systems.

%===============================================================================

\end{twocolumn}

\begin{onecolumn}
\begin{figure}
\begin{center}
\includegraphics[scale=0.38]{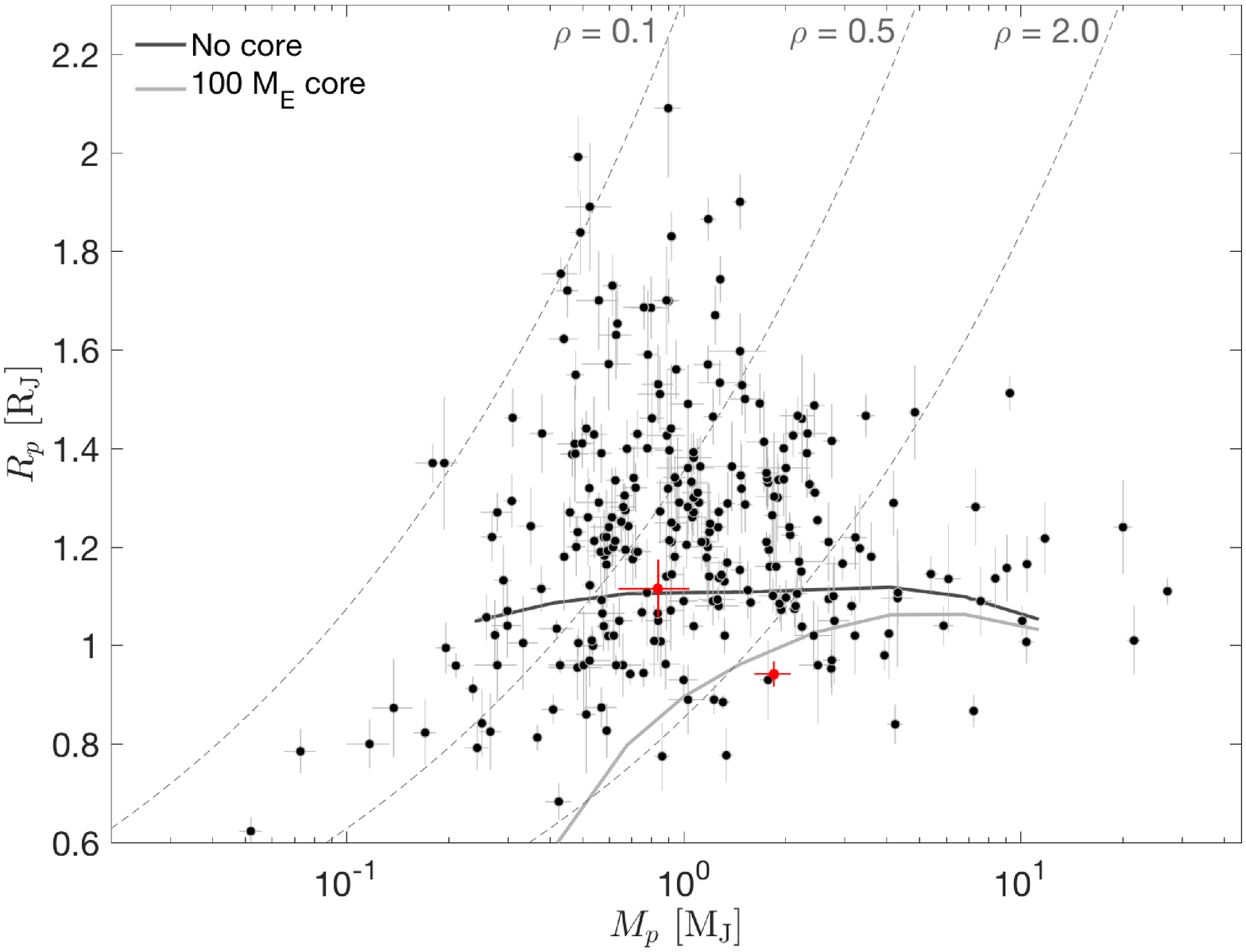}\\
\includegraphics[scale=0.38]{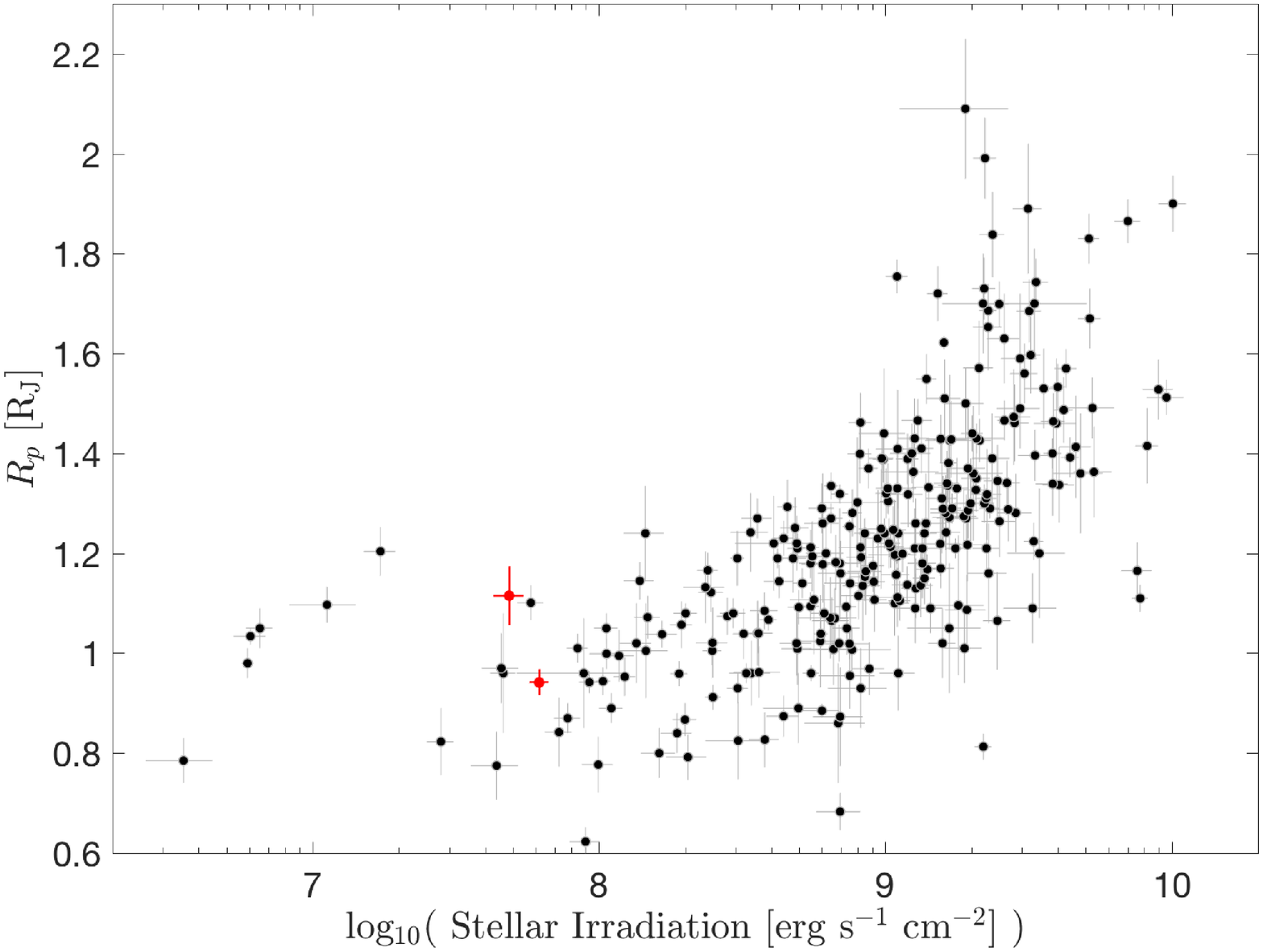}
\caption{Figures show the planet radius (linear scale) as a function of planet mass (log scale; top panel) and stellar irradiation (log scale; bottom panel). \tarob\ and \tartb\ are marked in red. In the top panel the solid lines encompass the theoretically expected region where planets are expected to reside in this parameter space \citep{fortney07}, between a core-less planet (solid black line) and a planet with a massive core of 100~\mearth\ (solid gray line). Both theoretical curves are for an assumed orbital star-planet separation of 0.045~au, although the planet radius changes by up to only $\approx$10\% between a separation of 0.02~au and 0.10~au \citep{fortney07}.
The dashed gray lines mark lines of equal mean density, and the density is labeled in the top part of the panel, in units of g~cm$^{-3}$.
Both panels show in black 273 planets with \rpl\ over 0.6 \rjup, planet radius error below 0.15 \rjup, and planet mass error below 20\% of the planet mass itself, to include only planets with well measured mass and radius. Circumbinary planets and directly imaged planets are excluded from these plots. Data shown in these plots was taken from the NASA Exoplanet Archive \citep{akeson13} on 2017 June 1st. 
\label{fig:radmassflux}}
\end{center}
\end{figure}
\end{onecolumn}

%===============================================================================
%===============================================================================

\begin{twocolumn}

\acknowledgments

% referee 
We are grateful to the anonymous referee for their meticulous reading of the manuscript and for providing detailed comments that helped improve this work. 
%Josh Pepper
We are grateful to Josh Pepper for his help in coordinating our ground-based follow-up observations with the KELT follow-up network.
% BJ Fulton
B.~J.~F.~acknowledges that this material is based upon work supported by the National Science Foundation Graduate Research Fellowship under Grant No.~2014184874.
% Andrew Vanderburg
A.~V.~is supported by the NSF Graduate Research Fellowship, grant No.~DGE 1144152.
% Diana Dragomir
D.~D.~acknowledges support provided  by NASA through Hubble Fellowship grant HST-HF2-51372.001-A awarded by the Space Telescope Science Institute, which is operated by the Association of Universities for Research in Astronomy, Inc., for NASA, under contract NAS5-26555.
% Andrew Cameron Collier 
A.~C.~C.~acknowledges support from STFC consolidated grant number ST/M001296/1.
% K2
This paper includes data collected by the \kt\ mission. Funding for the \kt\ mission is provided by the NASA Science Mission directorate.
% LCO
This work makes use of observations from the LCO network.
% Keck
Some of the data presented herein were obtained at the W.M. Keck Observatory, which is operated as a scientific partnership among the California Institute of Technology, the University of California and the National Aeronautics and Space Administration. The Observatory was made possible by the generous financial support of the W.M.~Keck Foundation.
The authors wish to recognize and acknowledge the very significant cultural role and reverence that the summit of Mauna Kea has always had within the indigenous Hawaiian community. We are most fortunate to have the opportunity to conduct observations from this mountain.

%===============================================================================
{\it Facilities:} 
\facility{Gemini:North (DSSI, NIRI)},
\facility{\kt}, 
\facility{Keck:I (HIRES)},
\facility{Keck:II (NIRC2)},
\facility{LCO (SBIG, Sinistro)},
\facility{Euler 1.2~m (CORALIE)}.

\end{twocolumn}

%===============================================================================
\appendix

\begin{twocolumn}

\section{\kt\ warm Jupiter transit candidate identified as a false positive}

In addition to the two \kt\ transiting warm Jupiters whose confirmation as planets was described above we have identified one \kt\ transiting warm Jupiter candidate, EPIC~212504617 ($P$=39.26~days), as a stellar binary, meaning it is a false positive. The \kt\ Campaign 6 phase folded transit light curve is shown in \figr{fplc}, derived in the same way as described in \secr{k2}.

\begin{figure}
\includegraphics[scale=0.4]{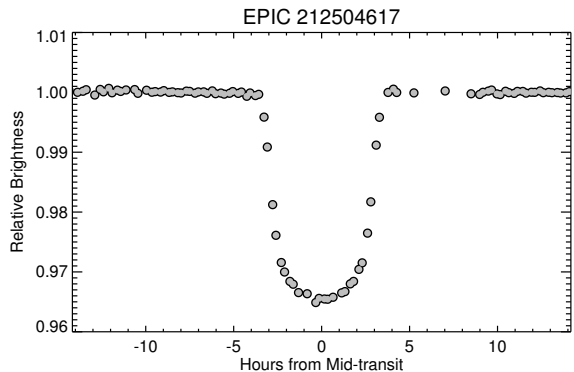}
\caption{\kt\ Campaign 6 phase folded transit light curve of EPIC~212504617, a warm Jupiter transiting candidate identified as a false positive. 
\label{fig:fplc}}
\end{figure}

% EPIC 212504617
We have identified EPIC~212504617 as a stellar binary using two RVs obtained with the CORALIE spectrograph, mounted on the Euler~1.2~m telescope in La Silla, Chile. CORALIE is a high resolution ($R$=60,000) fiber-fed echelle spectrograph that covers the wavelength range from 3900~\AA\ to 6800~\AA~\citep{queloz01}. Observations are made with a simultaneous Fabry-P\'{e}rot fiber to provide accurate wavelength calibration, and reduced via the standard CORALIE pipeline.  The stellar spectra are cross-correlated against a numerical mask with non-zero zones corresponding to stellar absorption features at zero velocity.

Two observations were made using CORALIE of the candidate EPIC~212504617, each with an exposure time of 2700~s. The two RVs have a difference of 16~\kms. Those RVs are listed in \tabr{fprv} and shown in \figr{fprvcurve}. The best-fit circular orbit model for these RVs gives a semi-amplitude of $K$=28.4~\kms\ and a systemic velocity of $\gamma=-15.5$~\kms. This is a two-parameter model fitted to only two RVs, hence this model is used only as an estimate for the orbital RV variation. Given the transit period of $P$=39.26~days and the host star's estimated mass of 1.01 \msun\ \citep{huber16}, the circular orbit RV semi-amplitude predicts a companion mass of about 0.6 \msun. Even when invoking a high eccentricity of 0.95 the companion mass should be at least 45 \mjup\ for the system to show an RV variability of 16 \kms. Therefore the companion cannot be a planet and is highly unlikely to be substellar.

\begin{deluxetable}{ccc}
\tablecaption{EPIC~212504617 Euler/CORALIE radial velocities \label{tab:fprv}}
\tablewidth{0pt}
\tablehead{
\colhead{Time} & \colhead{RV} & \colhead{RV err} \\
\colhead{BJD} & \colhead{\kms} & \colhead{\kms} 
}
\startdata
2457493.76167&	-24.667&	0.043\\
2457561.53544&	-40.603&	0.070\\
\enddata
\end{deluxetable}

\begin{figure}
\includegraphics[scale=0.42]{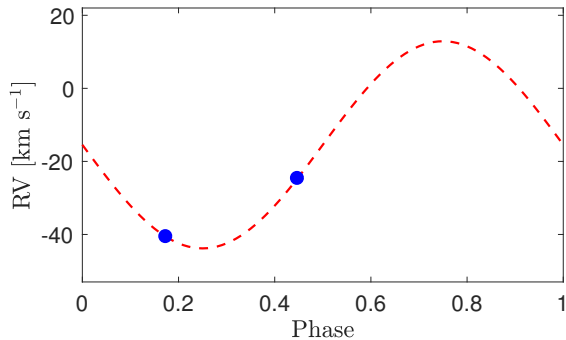}
\caption{Euler/CORALIE RVs of EPIC~212504617 phase folded at the transit period of $P$=39.26~day. RVs are marked in blue (error bars are smaller than the markers size), and the dashed line shows the best fit $e$=0 orbital solution. Phase zero is the primary eclipse (transit) phase. The fitted semi-amplitude is 28.4~\kms. Since this is a two-parameter model fitted to two RVs this model is used only as an estimate for the orbital RV variation. This estimate rules out the possibility of a sub-stellar mass companion, as discussed in more details in the text.
\label{fig:fprvcurve}}
\end{figure}

\end{twocolumn}

%===============================================================================

%===============================================================================

\end{document}